\documentclass[a4paper,11pt]{article}
\usepackage{pos}
\usepackage[export]{adjustbox}
\usepackage{subcaption}

\title{STACEX: RPC-based detector for a multi-messenger observatory in the Southern Hemisphere}
\ShortTitle{STACEX observatory}

\author*[a]{Rodriguez-Fernandez Gonzalo}
\author[b,f]{Bigongiari Ciro}
\author[c]{Bulgarelli Andrea}
\author[d,e]{Camarri Paolo}
\author[a]{Cardillo Martina}
\author[e]{Di Sciascio Giuseppe}
\author[c]{Fioretti Valentina}
\author[a] {Romani Marco}
\author[a]{Piano Giovanni}
\author[d,e]{Santonico Rinaldo}
\author[a]{Tavani Marco}

\affiliation[a]{IAPS-INAF\\
 Via del Fosso del Cavaliere 100, 00133, Rome, Italy}
\affiliation[b]{INAF-OAR,\\
 Via Frascati 33, 00078 Monteporzio Catone (Rome), Italy}
\affiliation[c]{INAF-OAS,\\
via Gobetti 101, I-40129 Bologna, Italy. }
\affiliation[d]{Dipartimento di Fisica Universit\'a di Roma Tor Vergata,\\
Via della Ricerca Scientifica 1, Rome, Italy}
\affiliation[e]{INFN - Roma Tor Vergata\\
Via della Ricerca Scientifica 1, Rome, Italy}
\affiliation[f]{ASI-Space Science Data Center\\
Via del Politecnico snc 00133, Rome, Italy}
\emailAdd{gonzalo.rodriguez@inaf.it}

 \abstract{Recent findings by the LHAASO experiment are opening a new window, that of the PeV sky, to the observation of the electromagnetic spectrum.
Several astronomical objects emitting gamma-rays at energies well above 100 TeV have been observed with the LHAASO-KM2 array of scintillators and muon detectors, clearly demonstrating the feasibility of gamma-ray astronomy up to PeV energies.
An all-sky gamma-ray detector in the Southern Hemisphere, operating in the GeV-PeV range, could complement LHAASO observations, monitor the Inner Galaxy and the Galactic Center looking for PeVatrons.
As shown by LHAASO, a water-Cherenkov based detector is not well suited to measure the energy spectrum up to the PeV range, nor to reach the advisable 100 GeV threshold.
The ARGO-YBJ experiment, operated for many years at 4300 m a.s.l. with an energy threshold of about 300 GeV, demonstrated, on the contrary, the capability of a carpet of Resistive Plate Chambers (RPCs) to fully reconstruct showers starting from the GeV range up to about 10 PeV.
In this contribution we propose a hybrid detector made of a layer of RPCs on top of a water Cherenkov facility devoted to the detection of muons for the selection of gamma-induced showers by the muon-poor technique.
We present the layout and discuss the expected performance.
}

\FullConference{37$^{\rm{th}}$ International Cosmic Ray Conference (ICRC 2021)\\
		July 12th -- 23rd, 2021\\
		Online -- Berlin, Germany}

\begin{document}
\maketitle

\section{Introduction}

The recent results obtained by the LHAASO Collaboration with half detector in data taking revealed the existence of a large number of gamma-ray sources emitting photons with energies well beyond 500 TeV \cite{lhaaso-nature}.
The unexpected observation of these sources in the Northern hemisphere suggests the opportunity to discover tens of similar Ultra High Energy emissions in the Inner Galaxy by a detector able to detect PeV photons located in the Southern hemisphere.

An ideal observatory for PeVatrons
\begin{itemize}
\item should be able to perform an unbiased survey to search for different and possibly unexpected classes of sources;
\item should have a dynamical range from 100 GeV to 10 PeV to measure the energy spectra with the same detector;
\item should have an effective area bigger than 0.5 km$^2$ to collect adequate photon statistics;
\item should have a very good energy resolution (20\% or better) above a few tens of TeV to test spectral break and cutoffs;
\item should have a good angular resolution ($\sim 0.2^{\circ}$) to resolve sources which might be hidden in the tails of bright sources and compare and correlate with gas surveys;
\item should have a background discrimination capability at a level of 10$^{-4}$ - 10$^{-5}$ starting from a few tens of TeV.
\end{itemize}

In this contribution we introduce the performance of a full coverage core detector (STACEX) constituted by
\begin{itemize}
\item[(a)] a 150$\times$150 m$^2$ RPC full coverage carpet, with a 0.5 mm lead layer on top;
\item[(b)] a dense muon detector array below the carpet constituted by water Cherenkov tanks LHAASO-like buried under 2.5 of soil;
\end{itemize}

Adequate photon statistics above 100 TeV can be provided by a dense scintillator and muon detector array LHAASO-like around this core detector covering a total area of 0.5 km$^2$ at least. 
Indeed, as shown by the LHAASO results, the performance of a water Cherenkov facility cannot compete with that of a scintillator array starting from a few tens of TeV. 
This layout is motivated by following reasons
\begin{itemize}
\item[(1)] dense sampling by the RPC carpet for a very low energy threshold ($\sim$100 GeV);
\item[(2)] wide energy range, 100 GeV $\to$ 10 PeV, with the same detector;
\item[(3)] high granularity of the carpet read-out to have an energy resolution $<$20\% above 10 TeV and to have a very good angular resolution ($\sim 0.2^{\circ}$ above 10 TeV);
\item[(4)] high efficiency rejection of the CR background by the muon-poor technique. With a 20,000 m$^2$ muon detector below the RPC carpet the background-free detection of gamma-rays is expected to start from a few tens TeV;
\item[(5)]  measurement of the elemental composition with two different independent observables, shower core characteristics (in a ARGO-like measurement) and muon component.
\end{itemize}

\section{Simulation}
\subsection{Simulation setup}

In this work we simulated a 150$\times$150 m$^2$ full coverage carpet constituted by bakelite-based RPCs, similar to the ARGO-YBJ experiment, located at 5000 m asl.
In fact, we successfully operated for a decade bakelite-based RPCs at 4300 m asl in the YangBaJing Cosmic Ray Laboratory in Tibet continuously flushing gas over an area of 110$\times$110 m$^2$. 
The detector has a modular structure, the basic data acquisition element being a cluster (5.7$\times$7.6 m$^2$), made of 12 RPCs (2.85$\times$1.23 m$^2$ each). Each chamber is read by 80 external strips of 6.75$\times$61.80 cm$^2$ (the spatial pixels), logically organized in 10 independent pads of 55.6$\times$61.8 cm$^2$ which represent the time pixels of the detector. 
The readout of 71277 pads and 570216 strips is the experimental output of the detector. In order to extend the dynamical range up to PeV energies, each chamber is equipped with two large size pads (139$\times$123 cm$^2$) to collect the total charge developed by the particles hitting the detector.

We have simulated gammas and protons primaries. The air shower development in the atmosphere has been generated with the CORSIKA\cite{corsika} version 77100. The selected hadronic interaction model is QGSJET01 for high energy and GHEISHA for low energy. The secondary particles have been propagated down to cut-off energies of 1 MeV (electromagnetic component) and 100 MeV (muons and hadrons). Primaries are simulated in the energy range from 10 GeV to 1 PeV with an index of -2.49 for gammas and -2.7 for protons. The zenith angle is fixed at 20$^\circ$ , the azimuth angle range from 0$^\circ$ to 360$^\circ$.
The experimental conditions (trigger logic, time resolution, electronic noises, relation between strip and pad multiplicity, etc.) have been taken into account using a GEANT4\cite{geant4} framework simulation. The core positions have been randomly sampled in an energy-dependent area large up to 600$\times$600 m$^2$, centered on the detector. 

We used a reconstruction procedure of the shower parameters similar to that used in ARGO-YBJ \cite{argo:moon} through the following steps.
At first, a plane surface is fitted (with weights equal to 1) to the shower front. This procedure is repeated up to 3 times, each iteration rejecting hits whose arrival time is farther than 2 standard deviations from the mean of the distribution of the time residuals from the fitted plane surface. This iterative procedure is able to reject definitively from the reconstruction the time values belonging to the non-Gaussian tails of the arrival time distributions. 
After this first step the problem is reduced to the nearly-vertical case by means of a projection which makes the fit plane overlapping the detector plane. 
Thereafter, the core position, i.e. the point where the shower axis intersects the detection plane, is obtained fitting the lateral density distribution of the secondary particles to a modified Nishimura-Kamata-Greisen (NKG)\cite{NKG} function:

\begin{equation}
\rho = \frac{{\rm N_{size}}}{2 \pi r^2_{M}}\frac{\Gamma(4.5-s)}{\Gamma(s)\Gamma(4.5-2s)}\left(\frac{r}{r_M}\right)^{(s-2)} \left(1+\frac{r}{r_M}\right)^{(s-4.5)}
\label{eq:nkg} 
\end{equation}

where $r$ is the distance to the core, N$_{{\rm size}}$ is the total number of particles, $s$ is the age of the shower, $r_M$ is the Moliere radius, and it is fixed to 133 m. The fit procedure is carried out via the maximum likelihood method using the MINUIT\cite{minuit} package.

Finally, the core position is assumed to be the apex of a conical surface to be fitted to the shower front. The slope of such a conical correction is fixed to $\alpha$ = 0.03 ns/m.

The analysis reported in this paper refers to events selected according to the following criteria: (1) more than 20 strips should be fired on the carpet; (2) the reconstructed core position should be inside an area 600$\times$600 m$^2$ centered on the detector. 

\subsection{Sensitivity estimation}

For the RPC array, we can estimate the number of detected gamma rays using:
\begin{equation}
N_\gamma = \int_E {\rm A}^\gamma_{{\rm eff}}(E) J_\gamma(E)dE \cdot {\rm T}_{{\rm obs}}
\end{equation}

and the number of background events according to the relation:
\begin{equation}
N_{{\rm cr}} = \int_E {\rm A}^{{\rm nuclei}}_{{\rm eff}}(E) J_{{\rm nuclei}}(E)dE \Delta\Omega(N_{{\rm trig}}) \cdot {\rm T}_{{\rm obs}}
\end{equation}

where $A^\gamma_{{\rm eff}}$ and $A^{{\rm nuclei}}_{{\rm eff}}$ are effective areas of the array for primary gamma rays and cosmic ray background respectively, $J_\gamma$\cite{crabspec} and $J_{{\rm nuclei}}$\cite{protonbkg} are differential spectrum of the gamma ray source and cosmic ray background respectively,  $\Delta\Omega$ is solid angle of observation, and T$_{{\rm obs}}$ is the observation time. The Crab Nebula flux is considered during the estimation of sensitivity of the array. The signiﬁcance is calculated using the formula:

\begin{equation}
S(\Delta N_{{\rm hits}}) = \frac{N_\gamma(\Delta N_{{\rm hits}})}{\sqrt{N_{{\rm bkg}}(\Delta N_{{\rm hits}})}} \cdot {{\rm Q}},
\end{equation}

where Q is the factor for $\gamma$/p discrimination:
\begin{equation}
{{\rm Q}} = \frac{{\rm Survival \, Ratio \, of \, \gamma}} {\sqrt{{\rm Survival \, Ratio \, of \, proton}}}
\end{equation}

\subsection{Performance of the RPC carpet}

\begin{figure}[t!]
\begin{minipage}[t]{.47\linewidth}
  \centerline{\includegraphics[width=\textwidth]{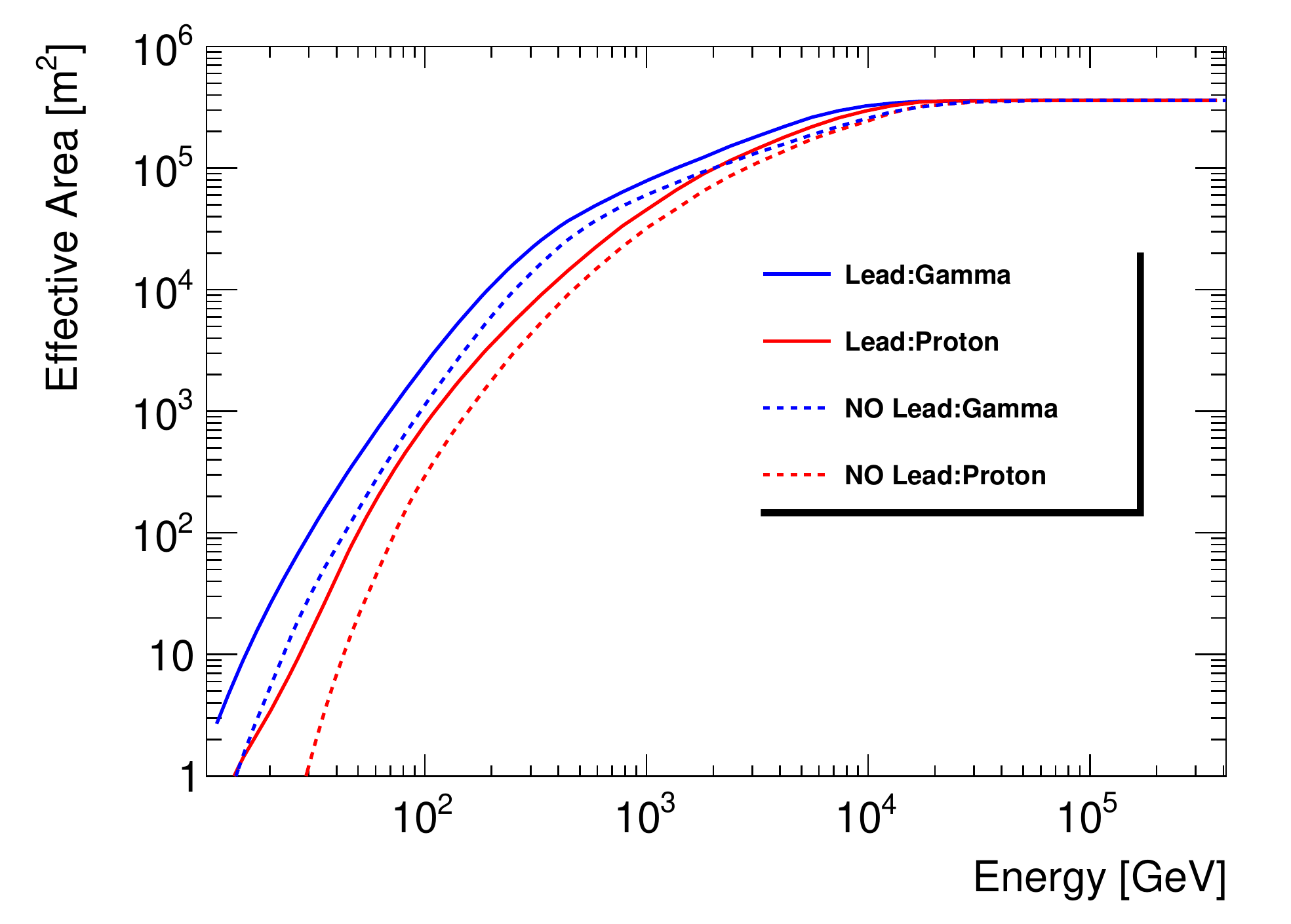} }
\caption[h]{STACEX effective areas for photon- and proton-induced showers as a function of the primary energy}
\label{fig:areaeff} 
\end{minipage}\hfill
\begin{minipage}[t]{.47\linewidth}
  \centerline{\includegraphics[width=\textwidth]{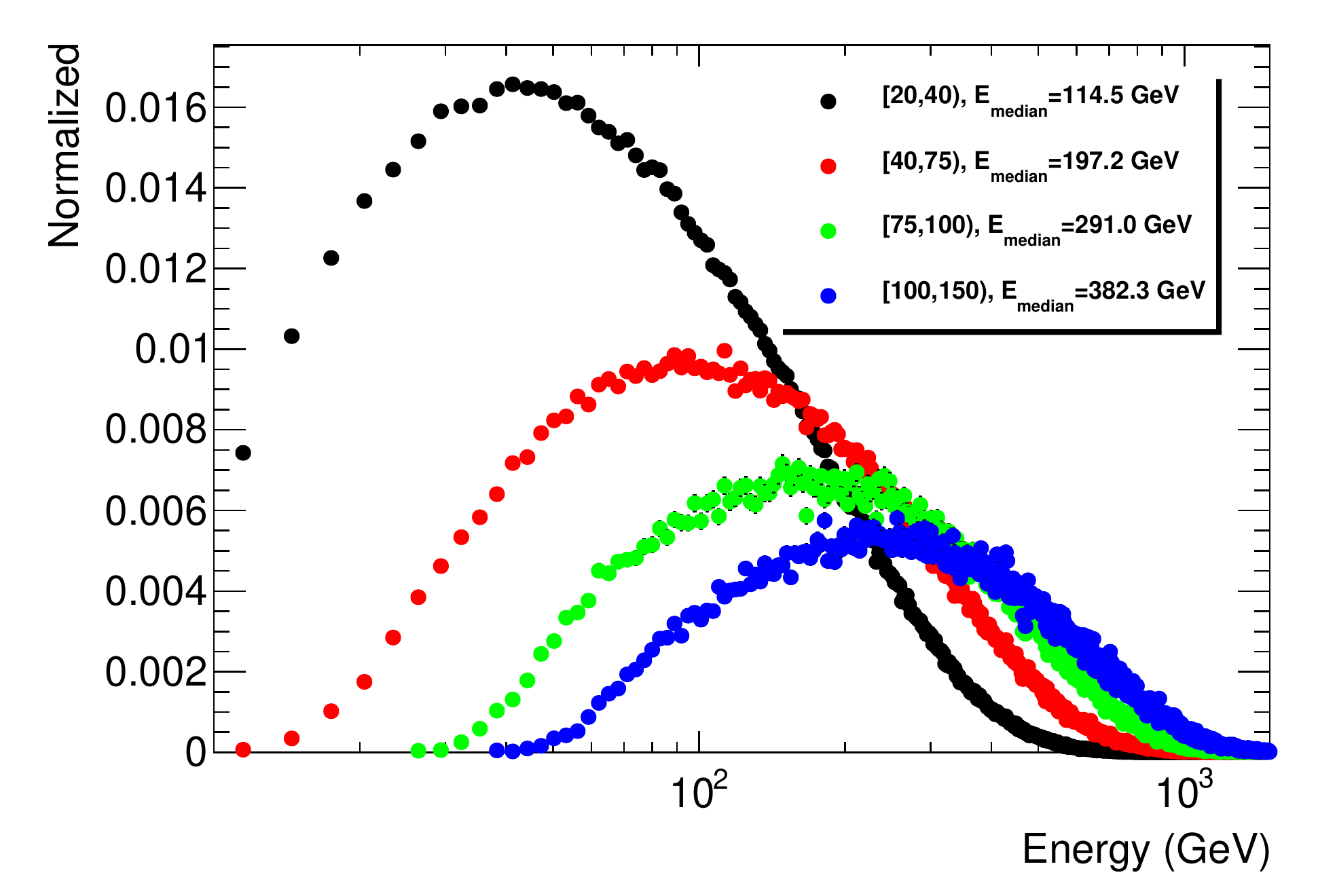} }
\caption[h]{Energy distributions of photon-induced showers for 4 different strip multiplicities measured by the RPC carpet.} 
\label{fig:medianene}
\end{minipage}\hfill
\end{figure}

\begin{figure}[t!]
\begin{minipage}[t]{.47\linewidth}
  \centerline{\includegraphics[width=\textwidth]{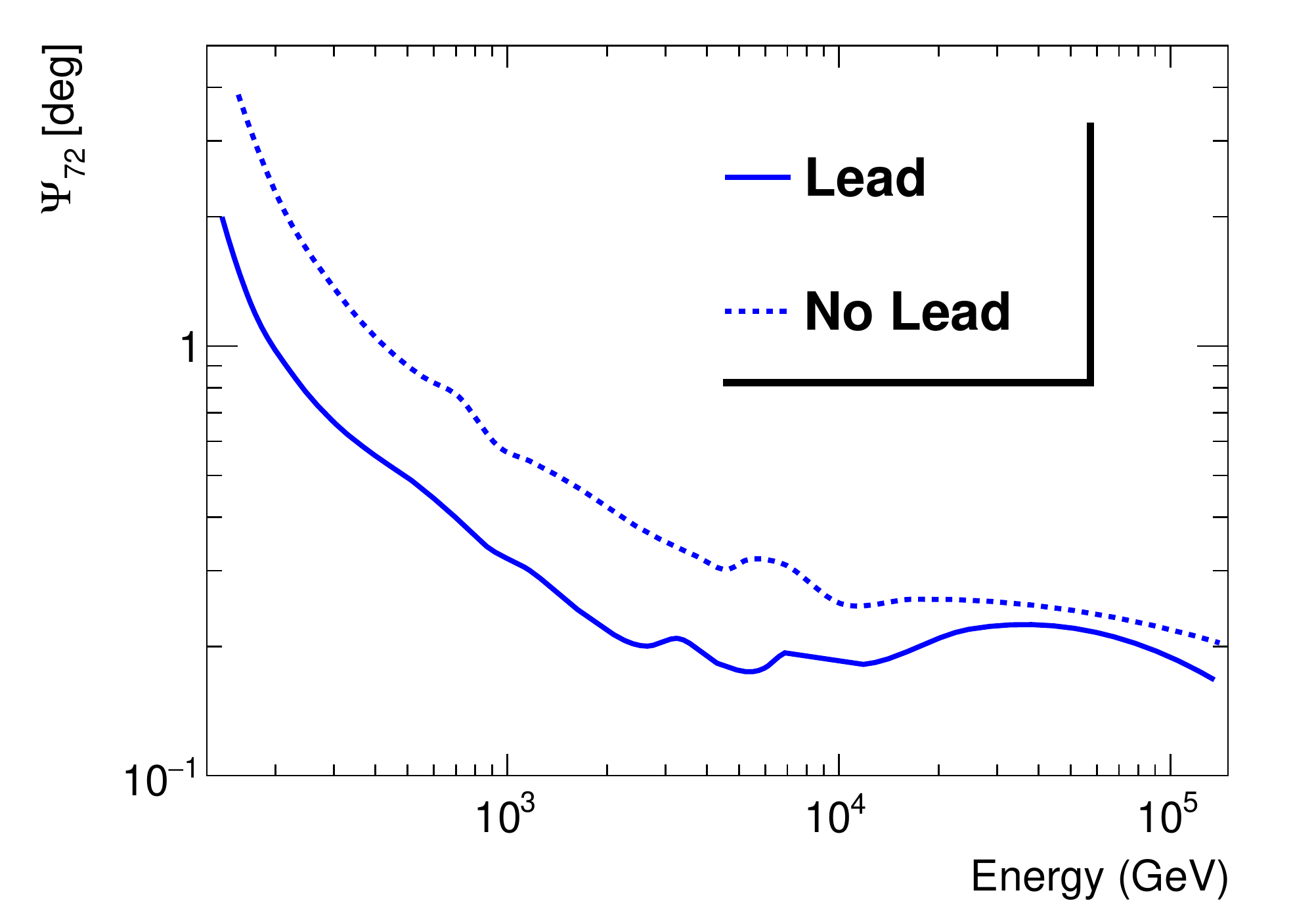} }
\caption[h]{Angular resolution for photon-induced showers as a function of the primary energy.}
\label{fig:angulresol} 
\end{minipage}\hfill
\begin{minipage}[t]{.47\linewidth}
  \centerline{\includegraphics[width=\textwidth]{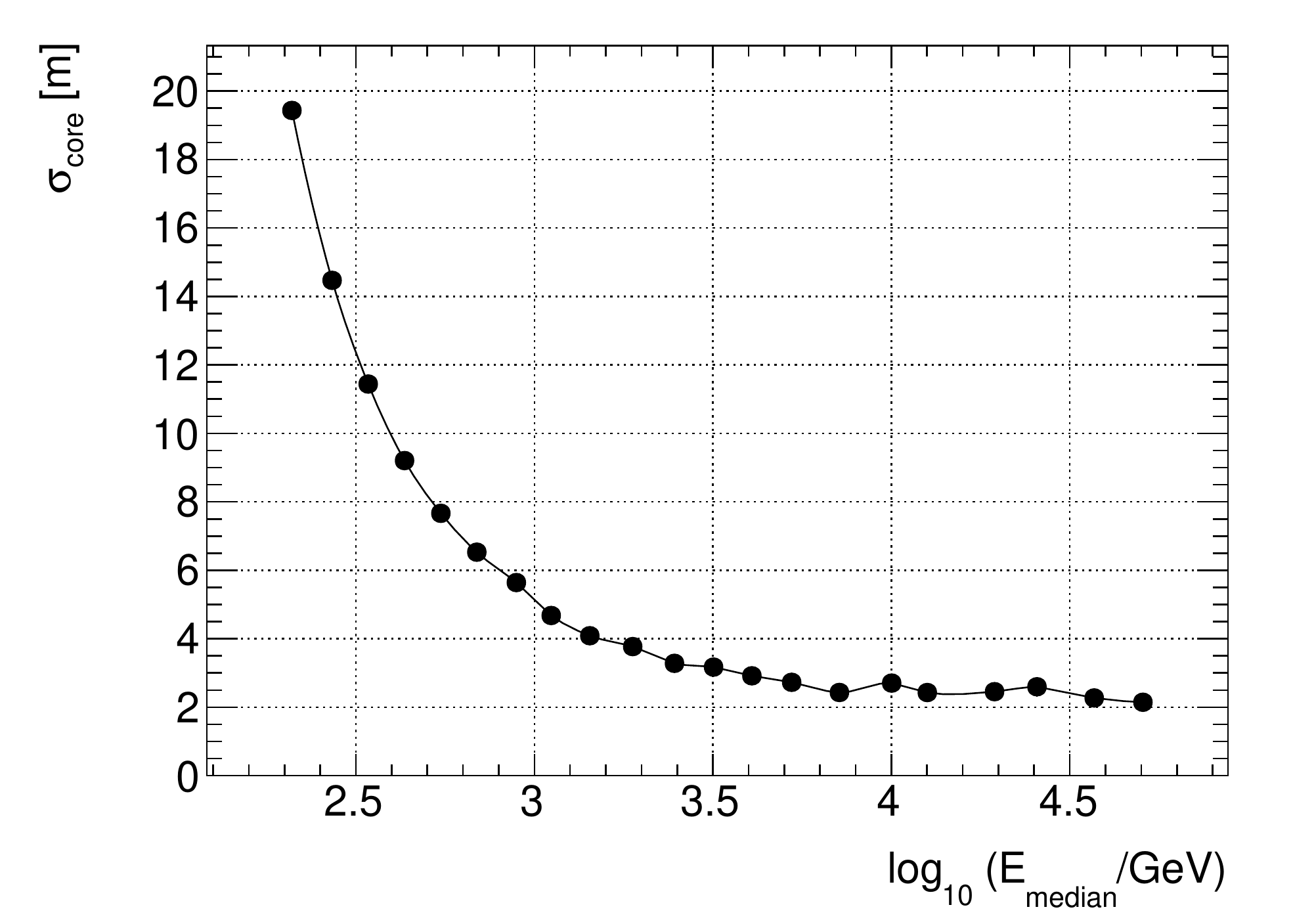} }
\caption[h]{Shower core resolution for photon-induced showers as a function of the primary energy.} 
\label{fig:coresol}
\end{minipage}\hfill
\end{figure}

The effective area for showers produced by primary photons and protons are shown in Fig. \ref{fig:areaeff} as a function of the median energy for different bins of strips multiplicity. As you can see from the figure, we have A$_\text{eff}\sim$ 3$\times$10$^3$ m$^2$ at 100 GeV and A$_\text{eff}\sim 10^6$ m$^2$ at 100 TeV. 
The energy distributions are shown in Fig. \ref{fig:medianene} for 4 different strip multiplicities. The peak energy of the first bin is about 100 GeV. The energy resolution is about 50$\%$.
In Fig. \ref{fig:angulresol} the angular resolution $\sigma_{\theta}$ (the angle containing the 72\% of the events) is shown. We have $\sigma_{\theta}\sim$0.5$^{\circ}$ at 1 TeV  and $\sigma_{\theta}\sim$0.25$^{\circ}$ at 10 TeV. The core resolution for gamma-ray events is shown in Figure \ref{fig:coresol} as a function of the reconstructed energy. The resolution is about 20 m at 100 GeV and $\sim$2 m at 100 TeV.

%
%

%
%

%
\begin{figure}[h]
\begin{subfigure}{0.33\textwidth}
\includegraphics[width=\textwidth]{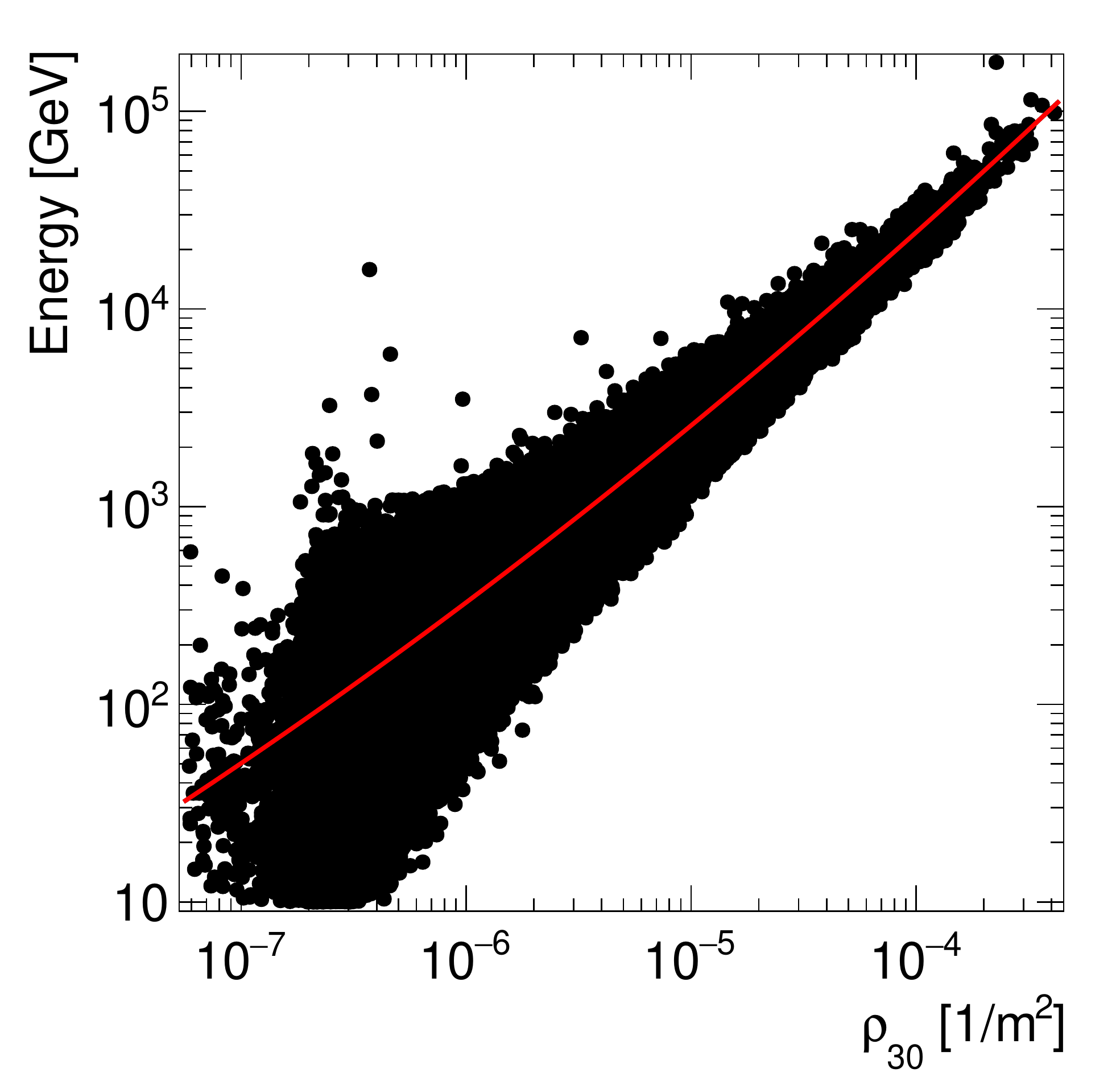} 
\end{subfigure}
\begin{subfigure}{0.33\textwidth}
\includegraphics[width=\textwidth]{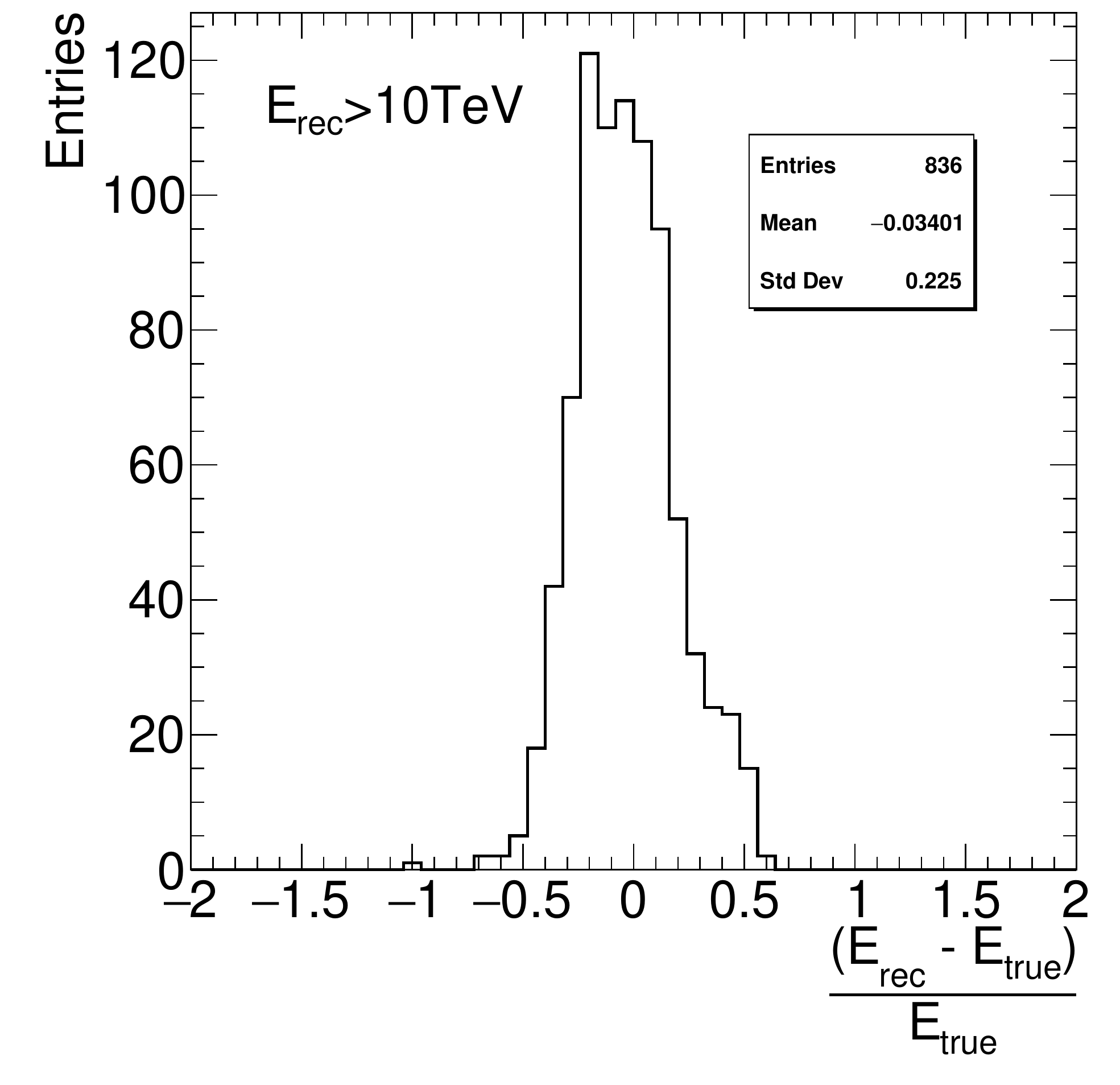} 
\end{subfigure}
\begin{subfigure}{0.33\textwidth}
\includegraphics[width=\textwidth]{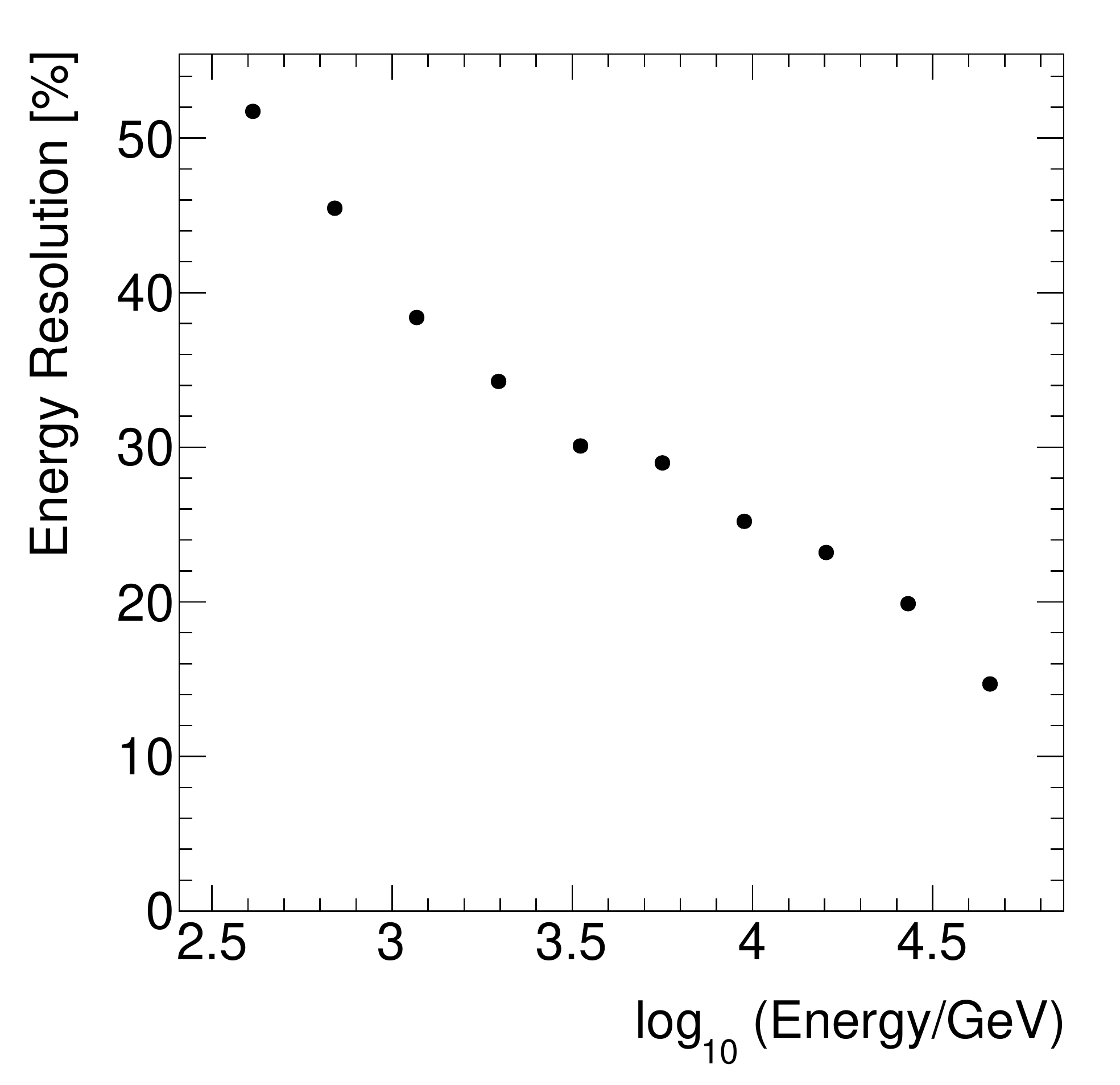} 
\label{fig:calplot} 
\end{subfigure}

\caption{The left panel gives the energy calibration using the $\rho_{30}$ estimator for simulated gamma-ray events for zenith angle of 20$^\circ$. The middle panel shows the energy resolution of showers in the energy range of 100 - 1000 TeV. The right panel shows the energy resolution vs the reconstructed energy.}
\label{fig:cali}
\end{figure}
%

\subsection{Energy reconstruction}

We have used the particle density at r=30m ($\rho_{30}$) evaluated using Equation \ref{eq:nkg} as a estimator of the gamma-ray energy.  The relation between $\rho_{30}$ and the primary energy is obtained using the equation:

\begin{equation}
\log(E_{{\rm rec}}/{\rm GeV} ) = {\rm a}\cdot (\log(\rho_{30}))^2 + {\rm b} \cdot \log(\rho_{30} ) + {\rm c}.
\end{equation}

In the left panel of figure \ref{fig:cali} the calibration plot is shown with the proposed fit. In the middle panel the energy resolution for $E_{{\rm rec}}>$ 10 TeV ($\sim$ 22 $\%$) is shown. Finally in the right panel we show the energy resolution as a function of the reconstructed energy. We note the the energy resolution at 100 TeV is $\sim$13$\%$.

\subsection{Discrimination of the CR background}

The classical method to reject the events induced by the background of charged CRs with arrays is to look for \emph{"muon poor"} showers. To evaluate the power of this background rejection technique it is important to know how frequently hadronic showers fluctuate in such a way to have a low muon content as the one resulting from $\gamma$-induced events \cite{disciascio-mu}. The main limitations of this technique is due to the extent of fluctuations in hadron-initiated showers and to the small number of muons. 
As shown by LHAASO, the capability to discriminate the CR background by the \emph{muon-poor technique} at a level of 10$^{-4}$ is the crucial point to study UHE gamma-ray astronomy with air shower arrays.

The basic elements are
\begin{itemize}
\item the total dimension of the muon detector;
\item the muon detection efficiency;
\item the coverage of the muon array.
\end{itemize}


%
%
%

\begin{figure}[h]
\begin{subfigure}{0.33\textwidth}
\includegraphics[width=\textwidth]{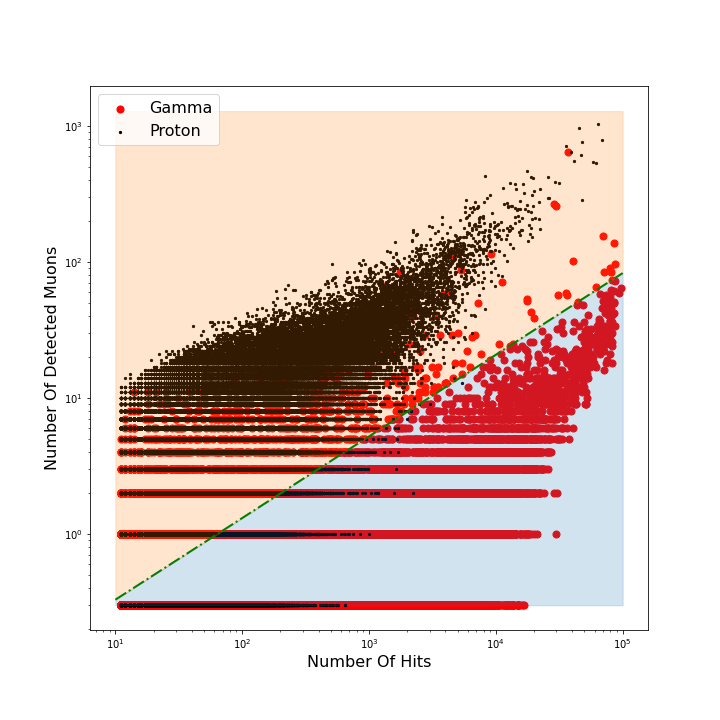} 
\end{subfigure}
\begin{subfigure}{0.33\textwidth}
\includegraphics[width=\textwidth]{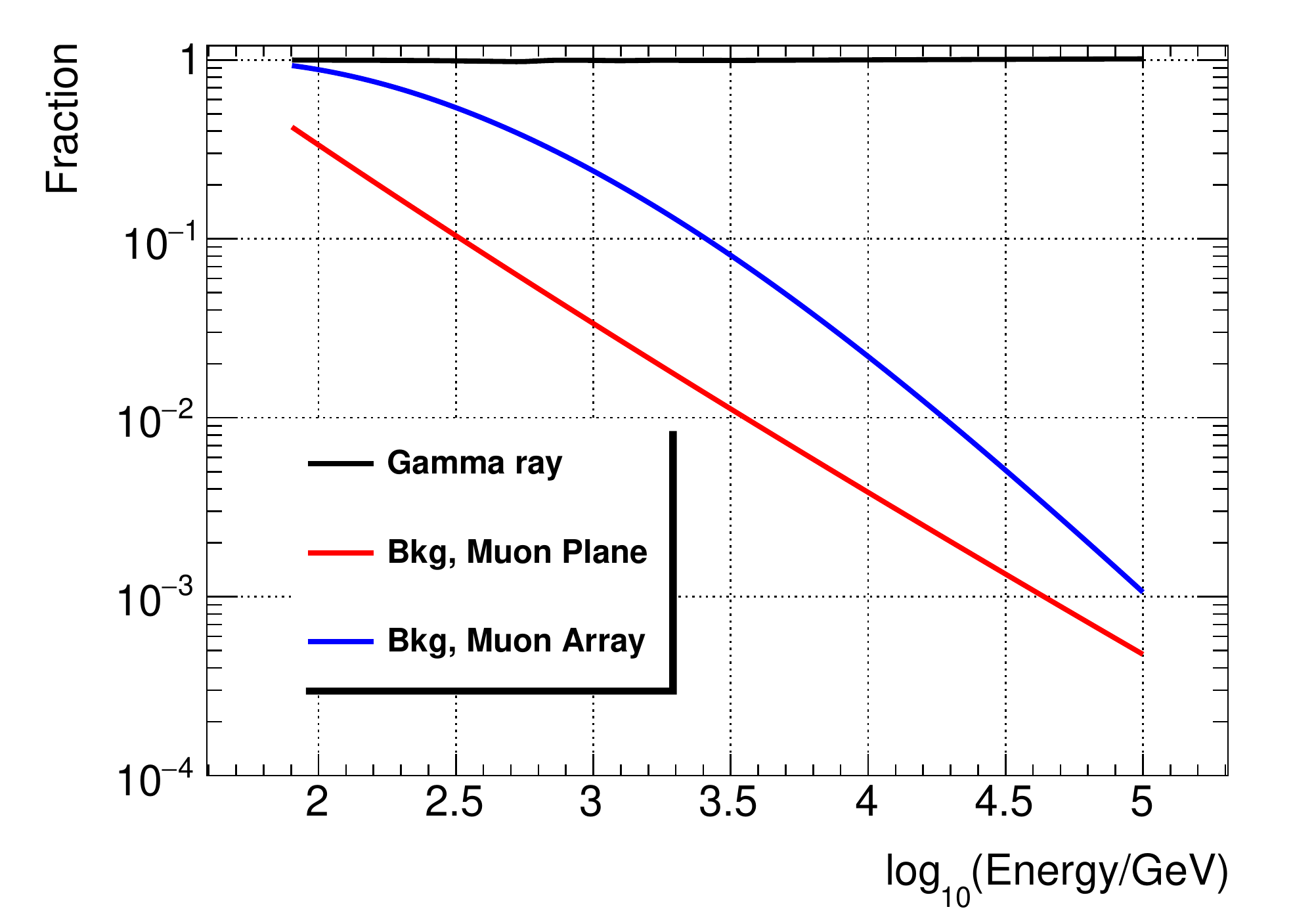} 
\end{subfigure}
\begin{subfigure}{0.33\textwidth}
\includegraphics[width=\textwidth]{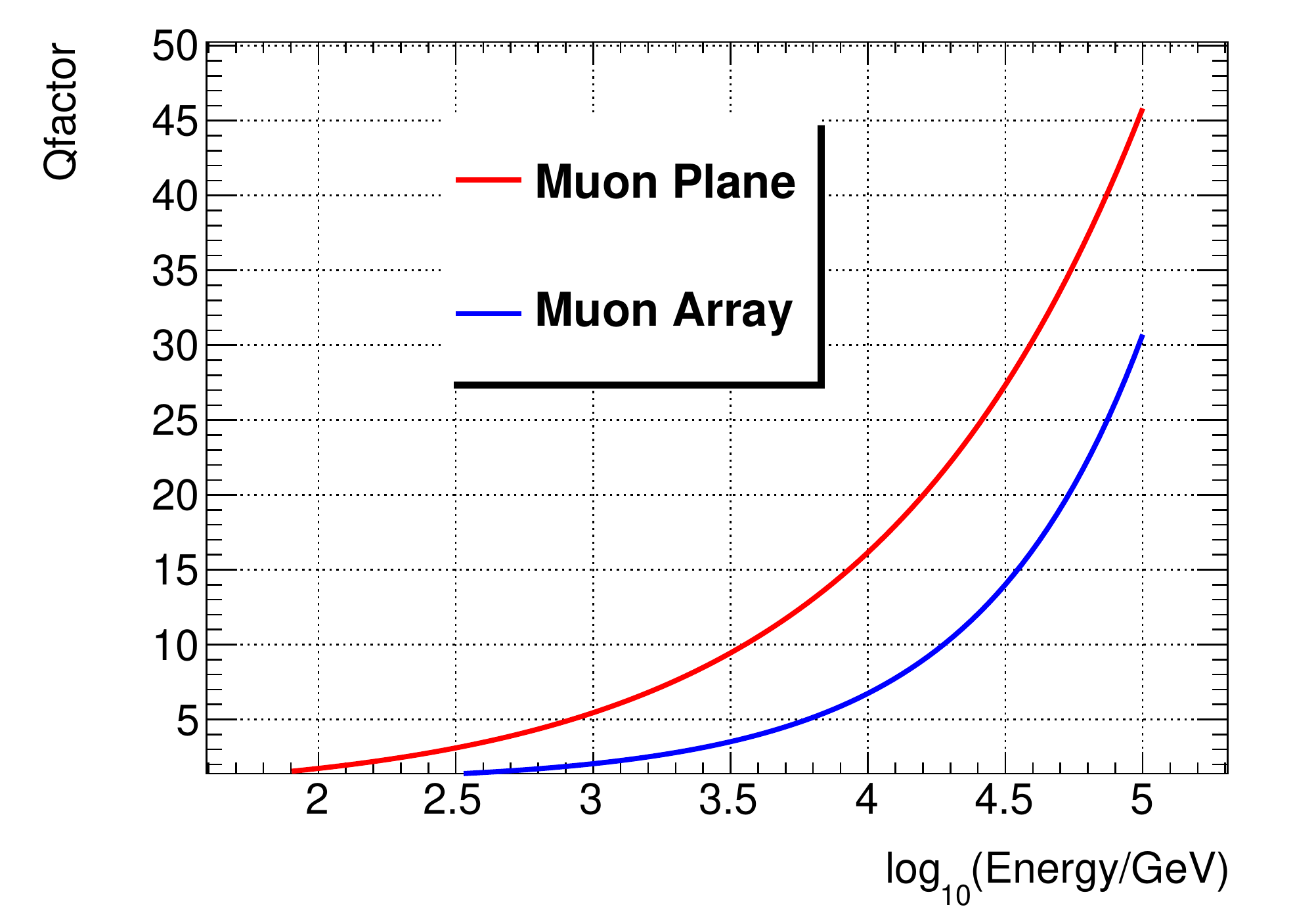} 
\end{subfigure}
\caption{Left panel: Number of muons detected by a 150$\times$150 m$^2$ continuous muon detector for proton and photon induced showers as a function of the strip multiplicity. Middel panel: Q-factor parameter as a function of the photon primary energy for the 2 investigated muon detector layouts. Right panel: The survival fraction of gamma-ray and cosmic ray background events as a function of the energy.}

\label{fig:discrimination}
\end{figure}

In this work we studied the CR background rejection using 2 different muon detector layouts: (1) a continuous muon detector below the RPC carpet with a total area of 22,500 m$^2$; (2) a 10$\times$10 array of LHAASO-like water tanks with a total area of 3,600 m$^2$.
In both cases the detectors are buried under 2.5 m of soil to reduce the punch-through probability by high energy secondary particles. As a consequence, the muon energy threshold is 1 GeV.
To further reduce the contamination in the analysis we exclude the muon detectors inside a circular area with 20 m radius around the reconstructed shower core position. 
In the left panel of Figure  \ref{fig:discrimination} the number of muons detected by a 150$\times$150 m$^2$ continuous muon detector for proton- and photon-induced showers is shown as a function of the strip multiplicity
We reject the CR background according to a selection cut removing showers with a muon content bigger than a value determined to optimize the sensitivity as a function of the multiplicity.
The fraction of the photons and protons surviving a selection cut determined using a binary classification method with a logistic function as a function of the number of strips is shown in the middle panel of Figure \ref{fig:discrimination}. For energies above 100 TeV  we reject the proton background at a level of 3$\cdot$ 10$^{-4}$ with nearly 100$\%$ of photons surviving. The so-called \emph{'Q-factor'} parameter is shown in the left panel as a function of the primary energy for the 2 investigated muon detector layouts. 

%
%
%

Calculations are still very preliminary but these results show that the background-free regime could start at a few tens of TeV.
%

\subsection{Sensitivity}
\begin{figure}[t!]
  \centerline{\includegraphics[width=\textwidth]{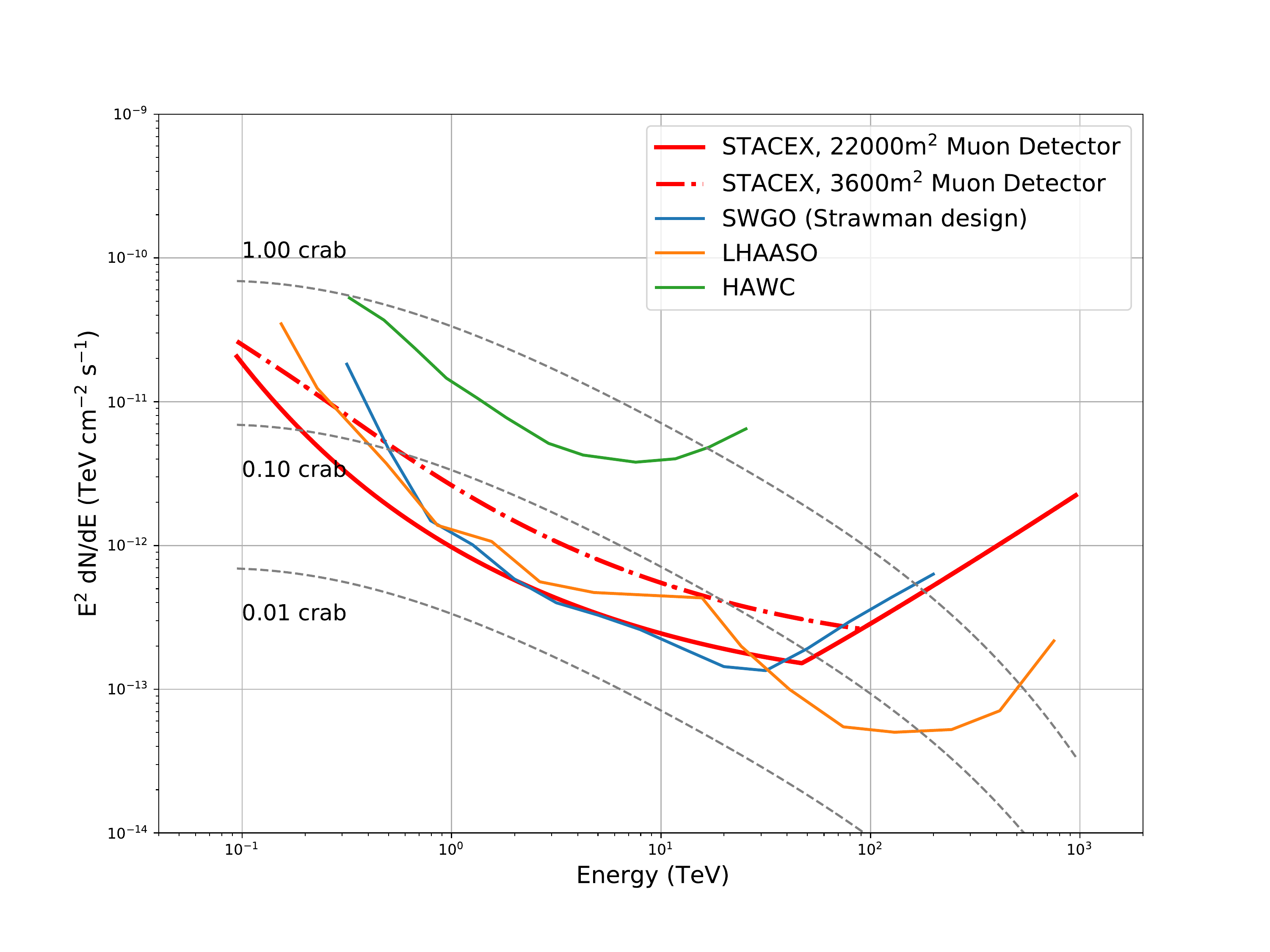} }
\caption[h]{Differential sensitivity of STATEX in 1 year compared to other experiments and projects. The band shows the range of values for the 2 layouts investigated in this work.}
\label{fig:sensitivity} 
\end{figure}

In Fig. \ref{fig:sensitivity} the sensitivity of STACEX is compared to that of other projects and experiments\cite{lhaaso, sgso, hawc}.
As discussed in this work, the crucial point in studying gamma-ray astronomy with air shower arrays is the background rejection via the muon-poor technique. We reported evidence that a 150$\times$150 m$^2$ hybrid array with a coverage of $\sim$90$\%$ could detect electromagnetic and muonic component with high efficiency reaching the sensitivity of larger infrastructures. The photon statistics to extend the energy range to 10 PeV can be provided by a dense array LHAASO-like.
%
%

\section{Conclusions}
The detection by LHAASO in the Northern hemisphere of a number of gamma sources emitting radiation well beyond 500 TeV opened the PeV sky to observations.
This result suggests that a much larger numbers of UHE sources are present in the Inner Galaxy and could be observed by a suitable  detector located in the Southern hemisphere.

In this work we showed that a hybrid detector made by a 150$\times$150 m$^2$ full coverage array able to image with high resolution both electromagnetic and muonic components could achieve these goals. 
The ARGO-YBJ Collaboration demonstrated that bakelite-based RPCs can be safely operated at extreme altitudes for many years providing: (1) high efficiency detection of low energy showers (energy threshold $\sim$100 GeV) by means of the dense sampling of the full coverage carpet; (2) unprecedented wide energy range investigated by means of the digital/charge read-outs ($\sim$100 GeV $\to$ 10 PeV); (3) good energy and angular resolutions with unprecedented details in the shower core region by means of the high granularity of the read-outs.
Coupled with a muon detector LHAASO-like this apparatus should be able to detect photons in a background-free regime starting from a few tens of TeV.

\end{document}